\documentclass[prb,aps,twocolumn,superscriptaddress,amsmath,amssymb]{revtex4-2}
\usepackage{amsmath}
\usepackage[english]{babel}
\usepackage{graphicx}
\usepackage{bm}
\usepackage{color}

\usepackage[colorlinks, linkcolor=red,citecolor=blue,urlcolor=blue]{hyperref}
\usepackage[all]{hypcap}

\begin{document}

\title{Proximity Induced Non-collinear Magnetic States in Planar Superconductor/Ferromagnet Hybrids}


\author{A. A. Kopasov}
\affiliation{National University of Science and Technology “MISIS”, Moscow 119049, Russia}
\affiliation{Institute for Physics of Microstructures, Russian Academy of Sciences, 603950 Nizhny Novgorod, GSP-105, Russia}
\author{S. V. Mironov}
\affiliation{Institute for Physics of Microstructures, Russian Academy of Sciences, 603950 Nizhny Novgorod, GSP-105, Russia}
\author{A. S. Mel'nikov}
\affiliation{Moscow Institute of Physics and Technology (National Research University), Dolgoprudnyi, Moscow region, 141701 Russia}
\affiliation{Institute for Physics of Microstructures, Russian Academy of Sciences, 603950 Nizhny Novgorod, GSP-105, Russia}

\begin{abstract}
The proximity induced superconducting (S) correlations in ferromagnetic (F) layers of planar S/F hybrids are shown to be responsible for the appearance of a nonlinear interaction between the magnetic moments of the F layers. This interaction originates from the combined influence of the orbital and exchange phenomena and can result in the spontaneous formation of non-collinear magnetic states in these systems. The proposed nonlinear coupling mechanism and resulting changes of magnetic textures are crucial for the design of the superconducting spintronics devices exploiting the long-range spin triplet proximity effect.
\end{abstract}

\maketitle

\section{Introduction}

During the past two decades planar hybrid structures consisting of superconducting (S) and ferromagnetic (F) layers were shown to host a huge variety of fascinating phenomena coming from the proximity effect, i.e. partial penetration of the Cooper pair wave function
from superconductors to ferromagnets. Among such phenomena one recognizes, e.g., the generation of spin-triplet superconducting correlations~\cite{Buzdin_rev, Golubov_rev, Bergeret_rev}, which has launched the rapid development of superconducting spintronics \cite{Eschrig_rev, Linder_rev, Melnikov_rev}, the formation of phase batteries including Josephson $\pi$ junctions~\cite{BuzdinJETPLett1982, RyazanovPRL2001, OboznovPRL2006} widely exploited in qubit engineering~\cite{FeofanovNP2010}, as well as  the spin-valve effect~\cite{OhAPL1997,TagirovPRL1999,BuzdinEPL1999} providing the tool for controllable switching of the sample resistance between zero and finite values at fixed temperature~\cite{Leksin, Zdravkov, Singh}.

The vast number of these theoretical and experimental studies are focused on the  thermodynamic, transport and magnetic properties of the superconducting state for a given magnetic ordering in ferromagnetic layers (see, e.g., Refs.~\cite{RobinsonS2010,KlosePRL2012,DiBernardoNC2015}). In most cases it is assumed that the magnetic texture is robust with respect to the superconducting phase transition, which allows one to engineer and analyze the desired magnetic configuration of ferromagnets in the non-superconducting state of the system and then assume this
configuration to remain the same at temperatures well below the superconducting critical temperature. Since the superconducting state is extremely sensitive to the magnetization profile in the F layers, this assumption appears to be a cornerstone of the correct interpretation of experimental data and design of devices for superconducting spintronics.
Thus, the understanding of possible mechanisms of the back-action, namely, the effect of superconducting ordering on the orientation of the magnetic moments in ferromagnets, is of primary importance for the progress in this field.

The physics underlying the back-action mechanisms can be related either to the orbital effects, i.e., to the changes in the stray fields of the magnetic subsystem due to the Meissner screening or to the exchange phenomena.
The stray fields penetrating the superconductor are known to induce the screening currents which, in their turn, induce magnetic fields piercing ferromagnets and affecting their magnetic state. This mechanism is responsible for the recently discussed helical instability in the array of magnetic particles placed on top of the superconductor~\cite{MukhamatchinJETPLett2011}, stabilization of magnetic skyrmions near the superconducting surface~\cite{VadimovAPL2018} etc. Note that for the structures including the F layers with the in-plane magnetization the above effects of the stray fields should be at first glance important only near the F film perimeter and, thus, negligible for the layers with the lateral sizes substantially exceeding their thicknesses. Interestingly, recently it was shown that the stray field related phenomena can be enhanced for the F film sandwiched between two superconductors~\cite{Mironov_Demagn}.

The influence of the exchange phenomena on the back-action of the superconducting order on the magnetic texture is known to be associated with the effects of the spin-triplet superconducting correlations. These correlations characterized by a finite spin polarization can also contribute to the magnetization and affect the magnetic state of the system~\cite{Krivorushko-PRB-02,Bergeret-PRB-03, Bergeret-EPL-04,Bergeret-PRB-04,Bergeret-PRB-05, Lofwander-PRL-05, Kharitonov-PRB-06,Faure-PhysC-07,Salikhov-PRL-09,Volkov-PRB-19}. The exchange interaction of the magnetic moments of the F subsystem with the spins of the electrons forming Cooper pairs can also result in modification of indirect interlayer exchange coupling of F layers. This phenomenon was predicted by P. de Gennes in his seminal paper~\cite{deGennesPL1966}, where he showed that the S film sandwiched between two ferromagnetic insulating (FI) layers can impose antiferromagnetic ordering of magnetizations of FI layers. Such superconducting exchange coupling, which has been recently observed experimentally in GdN/Nb/GdN trilayers~\cite{ZhuNM2017}, provides a possibility for control of the magnetic state in superconducting spintronics. Further theoretical studies of this problem (see, e.g., Refs.~\cite{SiprJPCM1995,SadeMeloPRL1996,BaladiePRB2003,GhanbariSR2021}) usually focus on collinear magnetic configurations and the question whether the proximity induced superconductivity can favor non-collinear magnetic configuration remains almost unexplored. Concerning the systems with the spin-orbit coupling, it has been recently shown that the non-aligned magnetic states can occur for coupled Josephson $\varphi_0$ junctions~\cite{BobkovPRB2022}, in which the Josephson energy is sensitive to the relative orientation of the exchange fields in the F layers.

\begin{figure}[t!]
	\includegraphics[width=0.37\textwidth]{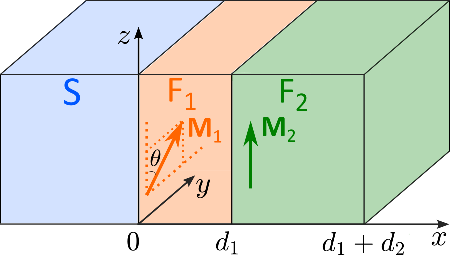}
	\caption{Sketch of the considered S/F$_1$/F$_2$ hybrid structures, in which the in-plane magnetic moments $\mathbf{M}_1$ and $\mathbf{M}_2$ form the angle $\theta$ with respect to each other. }
\label{Fig:Fig1}
\end{figure}

In this paper we demonstrate that the above mechanisms of the superconductor back-action on the magnetic moments in proximitized S/F structures with metallic ferromagnets can not be considered independently and the interplay between the orbital and exchange effects can result in a nontrivial nonlinear coupling of the magnetic moments giving rise to the temperature dependent reorientation transitions in the magnetic texture. Specifically, we study the above interplay for the exemplary layered S/F structure shown in Fig.~\ref{Fig:Fig1} and show that in S/F$_1$/F$_2$ trilayers the emerging superconductivity may favor the magnetic states with parallel, anti-parallel or even {\it non-collinear} orientation of magnetic moments in F$_1$ and F$_2$ layers. Thus, varying the temperature one may switch the magnetic subsystem between collinear and non-collinear configurations of magnetic moments. 
This finding is crucial for the superconducting spintronics since the S/F$_1$/F$_2$ trilayers serve as a basic building block for the devices exploiting the long-range spin triplet superconducting correlations.

The physics underlying the interplay mentioned above is associated with the strong dependence of the Meissner response of the induced superconducting correlations on the orientation of the magnetic moments.
Particularly important role  is played by the odd-frequency spin-triplet correlations providing the paramagnetic contribution to the electromagnetic response~\cite{Houzet_Meyer, Asano, Mironov_FFLO}. This paramagnetic contribution can suppress the total energy of screening currents making the non-collinear magnetic ordering more energetically favorable. Note that previously the importance of the contribution of the induced superconducting correlations to the electrodynamics of SF hybrids revealed itself in the so-called electromagnetic proximity effect~\cite{Mironov_APL, Devizorova, Putilov, FlokstraNP2016, FlokstraPRL2018}.

Let us also note that the interlayer exchange coupling in non-superconducting ferromagnetic multilayers has been the subject of active experimental and theoretical studies for several decades starting from seminal works~\cite{GrunbergPRL1986,HeinrichPRL1990,ParkinPRL1990, EdwardsPRL1991,BrunoPRL1991, BrunoPRB1995}. General form of the free energy contribution describing \textit{the interlayer coupling} for a ferromagnetic bilayer can be expressed as the expansion~\cite{BrunoPRB1995} $F(\theta) = \sum_{n = 0}^{\infty} J_n(\mathbf{M}_1\mathbf{M}_2)^n$, where $\mathbf{M}_1$, $\mathbf{M}_2$ are the magnetic moments of the layers and $\mathbf{M}_1\mathbf{M}_2 = M_1M_2\cos\theta$. The simplest nonlinear magnetic interaction beyond the bilinear Heisenberg-type interaction $J_1\mathbf{M}_1\mathbf{M}_2$ is the biquadratic exchange expressed by the contribution $J_2(\mathbf{M}_1\mathbf{M}_2)^2\propto \cos^2\theta$ (see also Ref.~\cite{DemokritovJPD1998}). We stress that the proximity induced reorientation transitions described in our work arise from the essentially nonlinear dependence of the free energy on $\mathbf{M}_1\mathbf{M}_2$ and can't be reduced just to some additional interaction symmetrically equivalent to the Dzyaloshinsky-Moria term (discussed, e.g., in~\cite{MukhamatchinJETPLett2011}). 
This nonlinear interaction results in the appearance of several free energy minima as a function of the relative angle $\theta$ and their competition with the change in temperature and system parameters. Certainly, for experimental observability of this effect the ferromagnetic bilayer should meet some important requirements: e.g., magnetization of one of the layers should be locked by the in-plane anisotropy due to proximity to the antiferromagnet while the in-plane magnetization anisotropy of another layer should be rather weak (see more detailed discussion in Sec.~\ref{results_sec}).

This paper is organized as follows. In Sec.~\ref{basic_eqs_sec} we present basic equations. Our main results are presented and discussed in Sec.~\ref{results_sec}. Finally, the results are summarized in Sec.~\ref{summary_sec}.

\section{Basic equations}\label{basic_eqs_sec}
To support the above qualitative picture we perform the microscopic calculations of the free energy of the system shown in Fig.~\ref{Fig:Fig1} as a function of the misorientation angle $\theta$ between the magnetic moments $\mathbf{M}_1$ and $\mathbf{M}_2$ of neighboring F layers of the thicknesses $d_{1}$ and $d_2$, respectively. We assume ${\bf M}_1$ and ${\bf M}_2$ to lay in the plane of the layers due to the easy-plane magnetic anisotropy. We choose the $x$-axis to be perpendicular to the layers and the $z$-axis to be directed along the magnetization ${\bf M}_2$, so that $\mathbf{M}_2 = M_2\mathbf{e}_z$ and $\mathbf{M}_1 = M_1\sin\theta\mathbf{e}_y + M_1\cos\theta\mathbf{e}_z$. For convenience we put the Planck constant ($\hbar$) and the Boltzmann constant ($k_B$) equal to unity. We assume that the system is in the dirty (Usadel) limit~\cite{Usadel}, which implies the local London-type relation between the superconducting current $\mathbf{j}_s$ and the vector potential $\mathbf{A}$. Let us note here that the spin-triplet superconducting correlations in the ferromagnet are odd-frequency correlations, have even spatial parity and are robust to the disorder scattering~\cite{Bergeret_rev}. The use of the Usadel formalism assumes some intermeduate values of the mean-free path: $l \ll \xi_0$ and $l\gg k_F^{-1}$, where $\xi_0$ is the superconducting coherence length and $k_F$ is the Fermi momentum.

Our analysis is based on the Usadel free-energy density, which reads as~\cite{EltschkaAPL2015,VirtanenPRB2020}
\begin{eqnarray}\label{Usadel_free_energy_general}
\mathcal{F}(\mathbf{r}) = \frac{\pi T \nu}{2}\sum_{\omega_n}{\rm Tr}\left\{[\check{\tau}_z (\omega_n + i\check{\Delta}) + i\check{M}]\check{g} + \frac{D}{4}(\check{\nabla}\check{g})^2\right\} \ .
\end{eqnarray}
Here $\omega_n = 2\pi T(n + 1/2)$ are the Matsubara frequencies ($n$ is an integer), $T$ is temperature, $\check{\nabla} = \nabla - i(e/c)[\mathbf{A}\check{\tau}_z, \cdot]$,  $c$ is the speed of light, $[\hat{\alpha},\hat{\beta}] = \hat{\alpha}\hat{\beta} -\hat{\beta}\hat{\alpha}$, $\check{\tau}_i$ ($i = x,y,z$) are the Pauli matrices acting in the particle - hole (Nambu) space, the system's Green's function $\check{g}$ has the following structure in the Nambu space
\begin{equation}
\check{g} = \begin{pmatrix}\hat{g}&\hat{f}\\ \hat{\bar{f}}&\hat{\bar{g}}\end{pmatrix} \ ,
\end{equation}
and obeys the normalization condition $\check{g}^2 = 1$,
\begin{equation}
\check{\Delta} = \begin{pmatrix}0&\Delta(i\hat{\sigma}_y)\\ \Delta^*(-i\hat{\sigma}_y)&0\end{pmatrix} \ , \ \ \check{M} = \begin{pmatrix}\mathbf{h}\hat{\boldsymbol{\sigma}}&0\\0&\mathbf{h}\hat{\boldsymbol{\sigma}}^*\end{pmatrix} \ ,
\end{equation}
$\hat{\sigma}_i$ ($i = x,y,z$) are the Pauli matrices acting in the spin space, $\Delta$ is the superconducting gap function, which is nonzero only in the S layer and assumed to be real, $\mathbf{h}$ is the exchange field, which is nonzero only in the F$_1$ and F$_2$ layers, $\nu$ is the density of states at the Fermi level, and $D$ is the diffusion coefficient. We assume the three latter quantities to be constant in each layer and, in particular, take the values $\mathbf{h}_j$, $\nu_j$, and $D_j$ in the F$_{j}$ layer ($j = 1,2$). The saddle point of the expression~(\ref{Usadel_free_energy_general}) yields the Usadel equations
\begin{equation}\label{Usadel_equations_nonlinear}
\check{\nabla}D\left(\check{g}\check{\nabla}\check{g}\right) + \left[\check{\tau}_z(\omega_n + i\check{\Delta}) + i\check{M},\check{g}\right] = 0 \ .
\end{equation} 
The spatial profile of the vector potential entering Eqs.~(\ref{Usadel_free_energy_general}) and (\ref{Usadel_equations_nonlinear}) is governed by the Maxwell equation
\begin{equation}\label{Maxwell_equation}
{\rm curl} \ {\rm curl} \mathbf{A} = (4\pi/c)\left(\mathbf{j}_m + \mathbf{j}_s\right) \ .
\end{equation}
Here $\mathbf{j}_m = c \ {\rm curl}\mathbf{M}$ is the magnetization current, and $\mathbf{j}_s(x) = - (c/4\pi)\lambda^{-2}(x)\mathbf{A}(x)$. We choose the following gauge for the vector potential: $\mathbf{A}(x) = A_y(x)\mathbf{e}_y + A_z(x)\mathbf{e}_z$. The inverse square of the London penetration depth $\lambda^{-2}(x)$, which also determines the spatial profile of the superfluid density $n_s(x)\propto \lambda^{-2}(x)$, is given by the standard expression (see, e.g., Ref.~\cite{Houzet_Meyer})
\begin{equation}\label{inverse_square_penetration_depth}
\lambda^{-2}(x) = [16\pi^2\sigma(x) T/c^2]\sum_{\omega_n > 0}\left(|f_s(x)|^2 - |\mathbf{f}_t(x)|^2\right) ,
\end{equation}
where $\sigma(x)$ is the normal-state conductivity, $f_s$ ($\mathbf{f}_t$) is the amplitude of the spin-singlet (spin-triplet) superconducting correlations, and $\hat{f} = (f_s + \mathbf{f}_t\hat{\boldsymbol{\sigma}})\hat{\sigma}_y$.

In what follows we make two key assumptions, which help us to simplify further calculations. First, considering the temperatures $T$ close to the superconducting critical temperature $T_c$, we can restrict ourselves to the free-energy contributions quadratic in the anomalous Green's function $\hat{f}$.  Second, we assume a large mismatch between the normal-state conductivities of the S ($\sigma_S$) and F$_{j}$ ($\sigma_{F_j}$) layers: $\sigma_S\gg \sigma_{F_{j}}$. In this limit the inverse proximity effect and the effects coming from the penetration of spin-triplet superconducting correlations to the S layer are small, so that the relevant $\theta$-dependent contributions to the free energy density stem only from the ferromagnetic layers. The resulting free-energy density of the structure can be written as a sum $\mathcal{F} = \mathcal{F}_{m} + \mathcal{F}_{ex} + \mathcal{F}_{gr}$ (see Appendix~\ref{App_A} for details of the derivation), where 
\begin{subequations}\label{free_energy_contributions}
\begin{align}
\label{magnetic_energy_density}
\mathcal{F}_m = \left({\rm curl}\mathbf{A} - 4\pi\mathbf{M}\right)^2/8\pi  + \lambda^{-2}\mathbf{A}^2/8\pi\ ,\\
\label{F_ex_def}
\mathcal{F}_{ex} = -2i\pi T \nu(x)\sum_{\omega_n > 0}\left[f_s\mathbf{h}(x)\mathbf{f}_t^* - f_s^*\mathbf{h}(x)\mathbf{f}_t\right] \ , \\
\label{F_gr_def}
\mathcal{F}_{gr} =   \pi T \nu(x) D(x)\sum_{\omega_n>0}\left[|\partial_x f_s|^2 - |\partial_x\mathbf{f}_t|^2\right] \ .
\end{align}
\end{subequations} 
In Eqs.~(\ref{free_energy_contributions}) the magnetic energy $\mathcal{F}_m$ is the sum of the magnetostatic energy and the energy of the screening currents, the term $\mathcal{F}_{ex}$ describes the exchange-type interaction between the exchange field and the spin polarization of the superconducting correlations, and $\mathcal{F}_{gr}$ is the contribution associated with the gradients of the Green's functions.

To find the spatial profiles of the anomalous Green's functions $f_s(x)$ and $\mathbf{f}_t(x)$ entering Eqs.~(\ref{inverse_square_penetration_depth}) and (\ref{free_energy_contributions}) we use the linearized Usadel equations. Inside the ferromagnets they take the form
\begin{subequations}\label{Usadel_equation}
\begin{align}
\left(D_{j}\nabla^2 - 2|\omega_n|\right) f_s = 2i \ {\rm sgn}(\omega_n)\mathbf{h}_j\mathbf{f}_t \ ,\\
\left(D_{j}\nabla^2 - 2|\omega_n|\right) \mathbf{f}_t = 2i \ {\rm sgn}(\omega_n)\mathbf{h}_j f_s \ . 
\end{align}
\end{subequations}
Due to the above assumption $\sigma_S\gg \sigma_{F_1}$ the function $\hat{f}$ satisfies the rigid boundary conditions at the S/F$_1$ interface (at $x = 0$), which gives $f_s(0) = \Delta/\sqrt{\omega_n^2 + \Delta^2}$ and $\mathbf{f}_t(0) = 0$. We put $\partial_x\hat{f} = 0$ at $x = d_1 + d_2$ and use the Kupriyanov-Lukichev boundary conditions at the F$_1$/F$_2$ interface (at $x = d_1$)
\begin{subequations}\label{KL_BC}
\begin{align}
\hat{f}(d_1 + 0) = \hat{f}(d_1 - 0) \ ,\\
\sigma_{F_2}\partial_x \hat{f}\biggl|_{d_1 + 0} = \sigma_{F_1}\partial_x \hat{f}\biggl|_{d_1 - 0} \ ,
\end{align}
\end{subequations}
disregarding, for simplicity, possible effects associated with a finite boundary resistance of the F$_1$/F$_2$ interface. The system~(\ref{Usadel_equation}) allows an analytic solution for an arbitrary relative angle $\theta$ between magnetic moments (the details of the derivation and rather cumbersome final expressions are provided in Appendix~\ref{Usadel_eq_solutions_app}).

\begin{figure*}[t!]
	\includegraphics[scale=0.67]{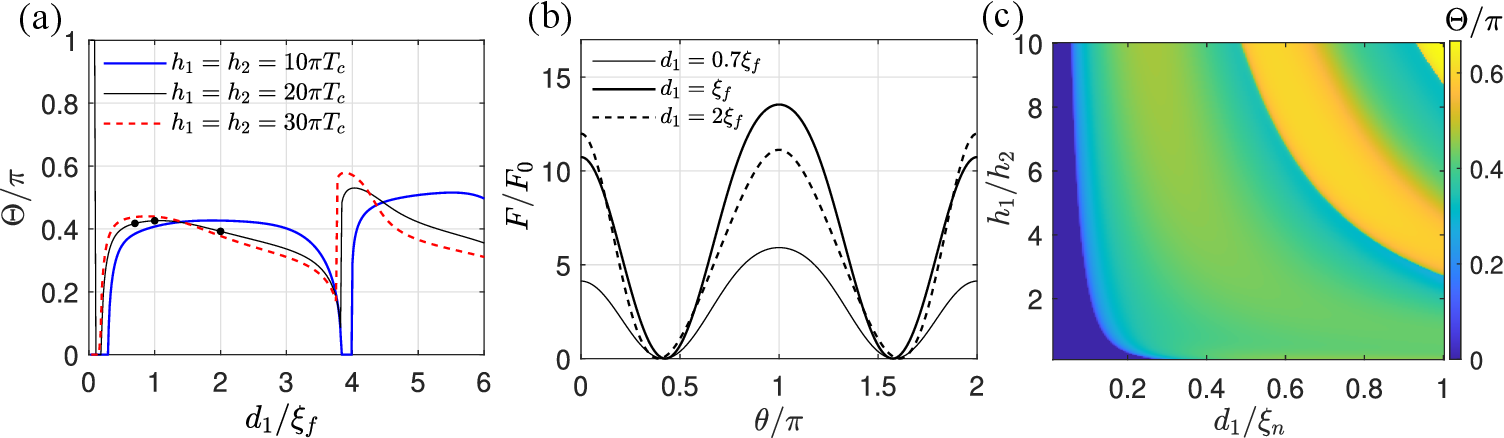}
	\caption{(a) Typical dependencies of the relative angle $\Theta$ between the magnetic moments in F$_1$ and F$_2$ layers corresponding to the the free-energy minimum on the thickness of the $F_1$ layer for $h_1 = h_2 = 10\pi T_c$, $20\pi T_c$, and $30\pi T_c$. Black dots highlight the parameters used in panel (b). (b) Free energy density per unit cross sectional area $F$ vs. the relative angle $\theta$ for $h_1 = h_2 = 20\pi T_c$ and $d_1/\xi_f = 0.7$, 1, and 2. (c) Colorplot of $\Theta$ as a function of $d_1$ and $h_1/h_2$ for $h_2 = 10\pi T_c$. Here $\xi_f = \sqrt{D/h}$, $\xi_n = \sqrt{D/2\pi T_c}$, $F_0 = (4\pi\alpha)^2\sigma\xi_n D\Delta^2/\mu_B^2c^2$, and we choose $\alpha k_Fl = 1$ and $d_2 = \xi_n$.}
\label{Fig:akfl1}
\end{figure*}

Rather simple analytical solution of the Maxwell equation~(\ref{Maxwell_equation}) can be obtained for a thick S layer with the thickness $d_s\gg \lambda$ and rather thin F layers with $d_1,d_2\ll \lambda$. The latter conditions allow us to neglect the spatial variations of the magnetic field ${\bf B} = {\rm curl}\mathbf{A}$ across the F layers. In the limit $\lambda_0/\lambda \to 0$ ($\sigma_{F_j}/\sigma_S \to 0$) and for zero applied magnetic field nonvanishing contributions to the magnetic energy per unit cross-sectional area $F_m = \int_{0}^{d_f}\mathcal{F}_m(x)dx$ originate from the second term in the right-hand side of Eq.~(\ref{magnetic_energy_density}) (see Appendix~\ref{App_C} for details of the derivation)
\begin{eqnarray}\label{magnetic_free_energy}
F_m(\theta)/2\pi = M_1^2\left[Q_2^{(1)}(\theta)+d_1^2Q_0^{(2)}(\theta)\right] \\
\nonumber
+2M_1M_2\cos(\theta)d_1Q_1^{(2)}(\theta) +M_2^2Q_2^{(2)}(\theta) \ ,
\end{eqnarray}
Here  $\lambda_0$ is the London penetration depth inside the superconductor, $d_f = d_1 + d_2$, $Q_n^{(1)}(\theta)=\int_{0}^{d_1}\lambda^{-2}(x,\theta)x^ndx$, and $Q_n^{(2)}(\theta) = \int_{d_1}^{d_f}\lambda^{-2}(x,\theta)(x-d_1)^ndx$. Although the free-energy contribution~(\ref{magnetic_free_energy}) describes the energy of the screening supercurrents and is of orbital origin, the exchange effects reveal themselves via the dependence of the local screening parameter $\lambda(x,\theta)$ on the relative angle $\theta$. The non-collinearity of magnetic moments gives rise to the formation of the long-range spin triplet correlations which make negative contributions to the screening parameter $\lambda^{-2}$ [see Eq.~(\ref{inverse_square_penetration_depth})] and, thus, to the magnetic kernels in Eq.~(\ref{magnetic_free_energy}), especially, $Q_2^{(2)}\left(\theta\right)$. In addition, they propagate over the distances $\sim\xi_n = \sqrt{D/2\pi T}$ from the F$_1$/F$_2$ interface which typically well exceed the decay scale $\sim\xi_f = \sqrt{D/h}$ for the spin-singlet ones. Thus, the presence of these long-range triplet correlations may result in the substantial decrease in the energy of screening supercurrents in Eq.~(\ref{magnetic_energy_density}) and make the non-collinear orientation of the magnetic moments energetically more favorable compared to the collinear one.

Let us note that the free-energy contributions $F_{ex} = \int_0^{d_f}dx \ \mathcal{F}_{ex}(x)$ and $F_{gr}= \int_0^{d_f}dx \ \mathcal{F}_{gr}(x)$ generally compete with the energy of the screening supercurrents and can hamper the formation of the non-collinear magnetic ordering. The interplay between the above mechanisms being sample specific depends on material parameters of ferromagnets and layer thicknesses. The key material parameter is the inverse coefficient of the molecular field $\alpha_j$ defined by the relation $\mu_B M_j = \alpha_j h_j $, where $\mu_B$ is the Bohr magneton. In calculations we parametrize the relative magnitude of the magnetic energy with respect to the exchange and the gradient terms by the parameter $(\alpha k_{F}l)_j$ relevant to the F$_j$ layer. Here $k_F$ is the Fermi wavenumber, and $l$ is the elastic mean-free path. Within the dirty limit quasiclassical description of superconductivity, which implies $k_Fl\gg 1$, the parameter  $\alpha k_Fl$ can vary within a wide range due to the fact that in a typical experimental situation $\alpha \ll 1$.

\begin{figure*}[t!]
	\includegraphics[width=0.95\textwidth]{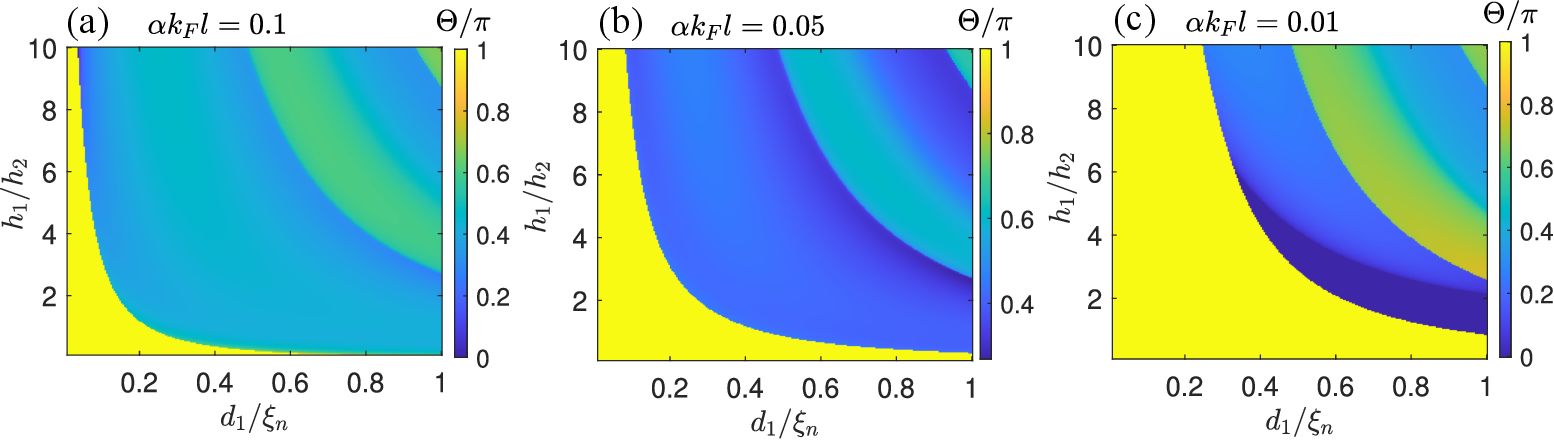}
	\caption{Typical colorplots of the equilibrium relative angle $\Theta$ versus $d_1$ and $h_1/h_2$ for various $\alpha k_Fl$, where $\alpha$ is the inverse coefficient of the molecular field, $k_F$ is the Fermi momentum, and $l$ is the mean-free path. We choose $d_2 = \xi_n$, $h_2 = 10\pi T$, $\alpha k_Fl = 0.1$ (a), 0.05 (b), and 0.01 (c).}
\label{Fig:small_akfl}
\end{figure*}

\section{Results and discussion}\label{results_sec}

In what follows we analyze the dependence of the ground state angle $\Theta$ between magnetic moments ${\bf M}_1$ and ${\bf M}_2$ on the key system parameters. Figs.~\ref{Fig:akfl1}(a) and \ref{Fig:akfl1}(c) show how $\Theta$ depends on the F$_1$ layer thickness $d_1$ for different ratios between the exchange field values in two ferromagnets. Fig.~\ref{Fig:akfl1}(b) reveals the behavior of the free energy density per unit cross sectional area as  a function of the relative angle $\theta$ for the parameters highlighted by black dots in Fig.~\ref{Fig:akfl1}(a). For our calculations we have chosen the case $\alpha k_Fl = 1$, for which we find that the electromagnetic proximity effect is rather strong for the broad range of parameters. Also, we choose equal diffusion constants, densities of states at the Fermi level and the inverse coefficients of the molecular field in both ferromagnets. Note that for $\Theta \neq \{0,\pi\}$ the magnetic states characterized by angles $\Theta$ and $(2\pi - \Theta)$ are necesserily degenerate, which follows from the discrete symmetry of the Usadel equations~(\ref{Usadel_equation}) $\hat{f}(\theta) = \hat{\sigma}_z\hat{f}(- \theta)\hat{\sigma}_z$. The results in Figs.~\ref{Fig:akfl1}(a) and \ref{Fig:akfl1}(c) refer to the branch $\Theta \in [0,\pi]$. Fig.~\ref{Fig:akfl1}(a) reveals highly non-monotonic $\Theta(d_1)$ dependencies and demonstrates a non-collinear magnetic ordering within wide parameter range. For rather small $d_1$, when the spin-triplet superconducting correlations in the system are not fully developed, the magnetic moments prefer a collinear alignment with $\Theta = 0$ or $\pi$. For sufficiently thick F$_1$ layer the collinear magnetic ordering becomes unstable and the system switches into a non-collinear state. The oscillatory $\Theta(d_1)$ behavior in Fig.~\ref{Fig:akfl1}(a) arises due to the oscillatory behavior of the Green's functions inside ferromagnetic layers and correlates with the amplitude of the spin-triplet components of the anomalous function at the F$_1$/F$_2$ interface. Regarding the results in Fig.~\ref{Fig:akfl1}(a) for $d_1\approx 4\xi_f$, we find that all the spin-triplet components of the anomalous function at the F$_1$/F$_2$ interface are rather small in comparison with the spin-singlet one. Our main results are summarized in Fig.~\ref{Fig:akfl1}(c), where we show the colorplot of the ground-state relative angle $\Theta$ as a function of the ratio $h_1/h_2$ and the thickness of the F$_1$ layer. 
Interestingly, the above described picture persists even for $\alpha k_Fl \lesssim 0.1$, for which the exchange and the gradient contributions to the free energy $F_{ex}$ and $F_{gr}$ can be rather significant. 
This fact is illustrated in Fig.~\ref{Fig:small_akfl}, where we show typical colorplots of the equilibrium relative angle $\Theta \in [0,\pi]$ between the magnetic moments for rather small $\alpha k_Fl$. In comparison with the results in Fig.~\ref{Fig:akfl1}(c) produced for $\alpha k_Fl = 1$, the colorplots in Figs.~\ref{Fig:small_akfl}(a), \ref{Fig:small_akfl}(b), and \ref{Fig:small_akfl}(c) have been obtained for $\alpha k_Fl = 0.1$, 0.05, and 0.01, respectively. The results in Fig.~\ref{Fig:small_akfl}(a) indicate that the magnetic moments favor a non-collinear alignment within wide ranges of the hybrid structure parameters even in the case $\alpha k_Fl \sim 0.1$. Further decrease in $\alpha k_Fl$ leads to an increase in the parameter range corresponding to a collinear orientation of the magnetic moments, which is demonstrated in Figs.~\ref{Fig:small_akfl}(b) and~\ref{Fig:small_akfl}(c).

Our results reveal the proximity induced nonlinear interaction between the magnetic moments in planar S/F hybrids. This interaction vanishes at the superconducting critical temperature $T_c$ and the interaction strength grows with decreasing temperature. In the vicinity of the critical temperature $F_m \propto \Delta^2 \propto (1- T/T_c)$. Our calculations show that the proposed nonlinear interaction should be more pronounced in structures characterized by $d_1\sim \xi_{f_1} = \sqrt{D_{F_1}/h_1}$, $d_2 \gtrsim \xi_{n_2} = \sqrt{D_{F_2}/2\pi T}$, and $h_1/h_2  > 1$. In particular, the results in Fig.~\ref{Fig:akfl1}(b) for $d_1 = \xi_f$ (shown by a thick black line) reveal $\delta F \equiv \max F(\theta) - \min F(\theta)\sim 10 F_0 \sim 10(4\pi)^2\xi_n\nu\Delta^2$. Taking, for instance, $\xi_n \sim 100$~nm, $\Delta \sim 0.1$~meV and typical density of states at the Fermi level $\nu \sim 10$~eV$^{-1}$nm$^{-3}$, we find $\delta F \sim 10^{-2}$~eV/nm$^2$. If the magnetic state is collinear at $T_c$, one can expect the emergence of temperature-driven transitions of the magnetic subsystem into a non-collinear state. Such transitions should reveal themselves in peculiarities of magnetization curves, which for a non-collinear magnetic ordering can be sensitive to the orientation of the in-plane external magnetic field. In addition, changing the orientation of the in-plane magnetic field one can also stimulate the transitions between degenerate non-collinear magnetic states accompanied by hysteresis phenomena. The formation of the non-collinear states driven by the long-range triplet correlations in S/F$_1$/F$_2$ hybrids with thick F$_2$ layer can be detected by measuring the local density of states (LDOS) at the Fermi level at the outer boundary of the ferromagnetic bilayer, i.e. at $x=d_f$. Indeed, if $d_2\gg\xi_f$ only spin triplet correlations should survive at $x=d_f$ which should result in the increase of LDOS as compared to the similar value in the normal state of the system~\cite{Braude, Buzdin_LDOS, Kontos, Cottet}.

Certainly, in typical experimental situations the suggested back-action mechanism of the proximity induced superconductivity on the magnetic texture should compete with various other interactions between the magnetic moments. Note that in this work we assumed the interfaces of the F layers to be smooth. At the same time the
interface roughness is known to produce stray fields leading to the so-called ``orange-peel'' interlayer coupling~\cite{NeelCRASF1962,SchragAPL2000,ChopraPRB2000,MoritzEPL2004,KuznetsovJMMM2019}. Experimental observation of our predictions also requires the nonlinear magnetic interaction to overcome the interlayer exchange coupling of F$_1$ and F$_2$ layers. An important point is that this interaction can be tuned and largely suppressed by including an additional normal metal (N) or insulating (I) layer between the F$_1$ and F$_2$ layers. This possibility arises due to oscillatory dependence of the interlayer exchange coupling on the thickness of the spacer. We expect that the conditions for observing the discussed nonlinear interaction should be more favorable in S/F$_1$/N/F$_2$ structures with an ultra-thin N layer than for S/F$_1$/I/F$_2$ systems due to fact that the inclusion of an insulating spacer can suppress the screening currents in F$_2$ layer. Note that the predicted reorinetation transition can be accompanied by the formation of the non-collinear domain structure resulting from the degeneracy of the states with misorientation angles $\Theta$ and $-\Theta$.

Finally, let us discuss the influence of the in-plane magnetic anisotropy in two ferromagnetic layers on experimental obervability of our predictions. First, in typical S/F$_1$/F$_2$ spin-valve structures the ferromagnetic materials are chosen to guarantee substantially different  magnetic anisotropy in the plane of the layers. Specifically, the direction of magnetic moment ${\bf M}_2$ is usually fixed by positioning of the adjacent antiferromagnet on top of the F$_2$ layer. In contrast, the F$_1$ layer is usually made of a soft ferromagnet (including ferromagnetic amorphous alloys and polycrystalline materials) with small magnetic anisotropy in the plane of the layer which makes it easy to switch the direction of ${\bf M}_1$ by relatively small external magnetic field. The magnetic hysteresis curves measured for typical superconducting spin-valves don't contain any signatures of significant in-plane magnetic anisotropy of the F$_1$ layer, so we expect that our model disregarding the in-plane magnetic anisotropy of the F$_1$ layer is relevant for experimentally realizable setups.

At the same time, characteristic energy scale associated with the effect discussed in our work can be comparable with typical energy barriers coming from the in-plane magnetic anisotropy of relevant ferromagnets. The typical anisotropy energy $E_a$ is usually in the range $10^2 - 10^7$~J m$^{-3}$~\cite{Blundell2001}. Taking, for instance, the thickness of the ferromagnetic layer of the order of 1 nm, we get that the anisotropy energy per unit cross-sectional area falls in the range $10^{-6} - 10^{-1}$ eV nm$^{-2}$. These values should be compared to the free-energy modulation originating from the energy of screening Meissner currents $\delta F = |\max F(\theta) - \min F(\theta)|$ ($\theta$ is the angle between the mangetic moments in ferromagnetic layers). Our estimate presented above $\delta F\sim 10^{-2}$~eV nm$^{-2}$ demonstrates that the nonlinear interaction due to superconducting correlations can be comparable with the in-plane anisotropy energy and the discussed effects can be rather significant in experiments even when the in-plane magnetic anisotropy in the F$_1$ layer is not negligible.

We expect that the S/F/F structures previously utilized for the study of the spin valve effect are good candidates to probe the spontaneous magnetic reorientation effects predicted in our paper. In particular, we expect that the predicted phenomena can occur in hybrid structures consisting of rather thick superconductor (for instance, Nb or Nb/Al alloy) in contact with two metallic ferromagnets with the in-plane magnetization (Fe, Ni or synthetic ferromagnet, such as Cu/Fe/Cu) [see, e.g., Refs.~\cite{Birge_1, Birge_2}]. Also we can expect that the proximity induced anisotropy axes in the plane of the F$_1$ layer should reveal themselves through the peculiarities of magnetization curves irrespective of the specific mechanism of magnetization reversal (domain nucleation, domain wall motion or coherent rotation).

\section{Summary}\label{summary_sec}
To sum up, we have shown that the interplay of the orbital and exchange effects in planar S/F hybrid structures results in the appearance of the nonlinear interaction between the magnetic moments of F layers. For dirty S/F$_1$/F$_2$ systems we have demonstrated that the proximity induced nonlinear interaction favors a non-collinear magnetic configuration in a broad parameter range typical for the experimentally fabricated structures.

\begin{acknowledgements}
The authors thank I. V. Bobkova, A. I. Buzdin, I. D. Tokman, and A. A. Fraerman for stimulating and useful discussions. This work was supported by the Russian Science Foundation (Grant No. 20-12-00053) in part related to the calculations of the magnetic energy and the Ministry of Science and Higher Education of the Russian Federation (Grant No. 075-15-2024-632) in part related to the analysis of the exchange effects. A. A. K. and S. V. M. acknowledge the financial support of the Foundation for the Advancement of Theoretical Physics and Mathematics BASIS (Grant No. 23-1-2-32).
\end{acknowledgements}

\appendix
\section{Derivation of Eqs.~(\ref{free_energy_contributions}) in the main text}\label{App_A}
In this section we briefly discuss the derivation of Eqs.~(\ref{free_energy_contributions}) in the main text. As a starting point, we write down the Usadel free energy density~(\ref{Usadel_free_energy_general}) in the ferromagnetic layers
\begin{equation}
\mathcal{F}(\mathbf{r}) = \frac{\pi T \nu}{2}\sum_{\omega_n}{\rm Tr}\left[(\omega_n\check{\tau}_z + i\check{M})\check{g} + \frac{D}{4}(\check{\nabla}\check{g})^2\right] \ .
\end{equation}
Here the trace is taken over the Nambu and spin indices. Introducing the functions $\hat{f}\to \hat{f}\hat{\sigma}_y$, $\hat{\bar{f}}\to\hat{\sigma}_y\hat{\bar{f}}$, and $\hat{\bar{g}}\to \hat{\sigma}_y\hat{\bar{g}}_y\hat{\sigma}_y$, we can rewrite the above expression in the following fashion
\begin{eqnarray}\label{free_energy_various_terms}
\mathcal{F}(\mathbf{r}) = \frac{\pi T\nu}{2}\sum_{\omega_n}{\rm Tr}\left[i\mathbf{h}\hat{\boldsymbol{\sigma}}(\hat{g}-\hat{\bar{g}})\right] + \\
\nonumber
 \frac{\pi T \nu D}{4}\sum_{\omega_n}{\rm Tr}\{(\nabla \hat{g})^2 + (\nabla\hat{\bar{g}})^2 +\\
 \nonumber
  2[(\nabla -2i(e/c)\mathbf{A}) \hat{f}][(\nabla + 2i(e/c)\mathbf{A}) \hat{\bar{f}}]\} \ .
\end{eqnarray} 
Considering the temperature $T$ close to the superconducting critical temperature, we substitute the expansions
\begin{subequations}
\begin{align}
\hat{g}\approx - {\rm sgn}(\omega_n)(1 - \frac{1}{2}\hat{f}\hat{\bar{f}}) \ ,\\
\hat{\bar{g}}\approx {\rm sgn}(\omega_n)(1 - \frac{1}{2}\hat{\bar{f}}\hat{f}) \ .
\end{align}
\end{subequations}
into Eq.~(\ref{free_energy_various_terms}) and restrict ourselves with the terms quadratic in the anomalous Green's functions. We get
\begin{eqnarray}\label{free_energy_preliminary}
\mathcal{F}(\mathbf{r}) = \frac{\pi T\nu}{4}\sum_{\omega_n}{\rm Tr}\left[i\mathbf{h}\hat{\boldsymbol{\sigma}}{\rm sgn}(\omega_n)(\hat{f}\hat{\bar{f}} + \hat{\bar{f}}\hat{f})\right] + \\
\nonumber
 \frac{\pi T \nu D}{2}\sum_{\omega_n}{\rm Tr}\{
  [(\nabla -2i(e/c)\mathbf{A}) \hat{f}][(\nabla + 2i(e/c)\mathbf{A}) \hat{\bar{f}}]\} \ .
\end{eqnarray} 
To cast the above expression into the form presented by Eqs.~(\ref{free_energy_contributions}) in the main text, we choose the gauge of the vector potential $\mathbf{A}(x) = A_y(x)\mathbf{e}_y + A_z(x)\mathbf{e}_z$. As a next step, we present $\hat{f} = f_s + \mathbf{f}_t\hat{\boldsymbol{\sigma}}$, $\hat{\bar{f}} = \bar{f}_s + \bar{\mathbf{f}}_t\hat{\boldsymbol{\sigma}}$ and use the symmetry relations $\bar{f}_s(\omega_n) = f_s^*(-\omega_n)$, $\hat{\bar{\mathbf{f}}}_t(\omega_n) = \mathbf{f}_t^*(-\omega_n)$, $f_s(\omega_n) = f_s(-\omega_n)$, and $\mathbf{f}_t(\omega_n) = -\mathbf{f}_t(-\omega_n)$, which follow from the Usadel equations. In particular, the matrix products entering the above free-energy contributions can be written as follows:
\begin{eqnarray}\label{f_matrix_products}
\hat{f}(\omega_n)\hat{\bar{f}}(\omega_n) = |f_s(\omega_n)|^2 - |\mathbf{f}_t(\omega_n)|^2 - \\
\nonumber
 f_s(\omega_n)\mathbf{f}_t^*(\omega_n)\hat{\boldsymbol{\sigma}} + f_s^*(\omega_n)\mathbf{f}_t(\omega_n)\hat{\boldsymbol{\sigma}} -  \\
 \nonumber
 i\left(\mathbf{f}_t(\omega_n)\times\mathbf{f}_t^*(\omega_n)\right)\hat{\boldsymbol{\sigma}} \ .
\end{eqnarray}
Substituting Eq.~(\ref{f_matrix_products}) into Eq.~(\ref{free_energy_preliminary}), we obtain Eqs.~(\ref{free_energy_contributions}) in the main text.

\section{Solution of the linearized Usadel equations~(\ref{Usadel_equation})}\label{Usadel_eq_solutions_app}
Here we provide an analytical solution of the linearized Usadel equations~(\ref{Usadel_equation}). The anomalous Green's function in the F$_2$ layer (at $d_1< x <d_1 + d_2$) $\hat{f} = f_s + f_{tz}\hat{\sigma}_z + f_{ty}\hat{\sigma}_y$ satisfying the condition $\partial\hat{f}/\partial x = 0$ at the F$_2$/vacuum interface ($x = d_f$), can be presented in the form
\begin{subequations}\label{second_layer_solutions}
\begin{align}
f_s = 2{\rm Re}\left[c_+\cosh[p_2(x-d_f)]\right] \ ,\\
f_{t_z} = 2i \ {\rm sgn}(\omega_n){\rm Im}\left[c_+\cosh[p_2(x-d_f)\right] \ ,\\
f_{t_y} = c \ {\rm sgn}(\omega_n)\cosh[q_2(x - d_1 - d_2)] \ .
\end{align}
\end{subequations}
Here $c_{+}$ and $c$ are the matching constants, and 
\begin{equation}
p_j = \sqrt{\frac{2\left(|\omega_n| + ih_j\right)}{D_{F_j}}} \ , \ \ q_j = \sqrt{\frac{2|\omega_n|}{D_{F_j}}} \ ,
\end{equation}
$j = 1$ and 2 for the F$_1$ and F$_2$ layer, respectively. Considering the linearized Usadel equations in the F$_1$ layer and introducing the transformed functions $\hat{\tilde{f}} = \hat{U}^{\dagger}\hat{f}\hat{U}$ and $\hat{\tilde{f}} = \tilde{f}_s + \tilde{\mathbf{f}}_t\hat{\boldsymbol{\sigma}}$ with $\hat{U}(\theta) = \cos(\theta/2) + i\hat{\sigma}_x\sin(\theta/2)$, we get the following system of equations 
\begin{subequations}
\begin{align}
D_{F_1}\nabla^2\tilde{f}_s - 2|\omega_n|\tilde{f}_s = 2ih_1{\rm sgn}(\omega_n)\tilde{f}_{t_z} \ ,\\
D_{F_1}\nabla^2\tilde{f}_{t_z} - 2|\omega_n|\tilde{f}_{t_z} = 2ih_1{\rm sgn}(\omega_n)\tilde{f}_s \ ,\\
D_{F_1}\nabla^2\tilde{f}_{t_y} - 2|\omega_n|\tilde{f}_{t_y} = 0 \ .
\end{align}
\end{subequations}
Correspondence between the initial and transformed functions is as follows:
\begin{subequations}\label{correspondence}
\begin{align}
f_s = \tilde{f}_s \ ,\\
f_{t_y} = \tilde{f}_{t_y}\cos(\theta)+\tilde{f}_{t_z}\sin(\theta) \ ,\\
f_{t_z} = -\tilde{f}_{t_y}\sin(\theta) + \tilde{f}_{t_z}\cos(\theta) \ .
\end{align}
\end{subequations}
Neglecting the inverse proximity effect in the S layer and imposing the rigid boundary conditions $f_s = f_{s0} = \Delta/\sqrt{\omega_n^2 + \Delta^2}$ and $\mathbf{f}_t = 0$ at the S/F$_1$ interface (at $x = 0$), we write down the general solution in the F$_1$ layer
\begin{subequations}\label{first_layer_solutions}
\begin{align}
\tilde{f}_s = {\rm Re}\left[f_{s0}\cosh(p_1x) + 2a\sinh(p_1x)\right] \ ,\\
\tilde{f}_{t_z} = {\rm sgn}(\omega_n)i{\rm Im}\left[f_{s0}\cosh(p_1x)+2a\sinh(p_1x)\right] \ ,\\
\tilde{f}_{t_y} = {\rm sgn}(\omega_n)b \sinh(q_1x) \ ,
\end{align}
\end{subequations}
where $a$ and $b$ are the matching constants. Matching the solutions~(\ref{second_layer_solutions}) and (\ref{correspondence}), (\ref{first_layer_solutions}) at the F$_1$/F$_2$ interface with the help of the Kupriyanov - Lukichev boundary conditions~(\ref{KL_BC}), we obtain the linear system of equations for the coefficients
\begin{subequations}
\begin{align}
{\rm Re}(c_+\mu_2)  = {\rm Re}(f_{\mu \nu}) \ ,\\
c\eta_2 = \cos(\theta)b\zeta_1 + 2i\sin(\theta){\rm Im}\left(f_{\mu\nu}\right) \ ,\\
{\rm Im}(c_+\mu_2)  = i\sin(\theta)b\zeta_1/2 + 
 \cos(\theta){\rm Im}\left(f_{\mu\nu}\right) \ ,\\
{\rm Re}(p_1f_{\nu\mu}) = 
 -(\sigma_{F_2}/\sigma_{F_1}){\rm Re}\left(c_+p_2\nu_2) \right) \ ,\\
\cos(\theta)bq_1\eta_1 + 2i\sin(\theta){\rm Im}(p_1 f_{\nu\mu}) = \\
\nonumber
 -(\sigma_{F_2}/\sigma_{F_1})cq_2\zeta_2 \ ,\\
-\sin(\theta)bq_1\eta_1 + 2i\cos(\theta){\rm Im}(p_1f_{\nu\mu}) = \\
\nonumber
 -(\sigma_{F_2}/\sigma_{F_1})2i{\rm Im}(c_+p_2\nu_2)   \ .
\end{align}
\end{subequations} 
Here $\sigma_{F_1}$ and $\sigma_{F_2}$ denote the normal-state conductivity in the F$_1$ and F$_2$ layer, respectively. To compactify the relations we introduced $f_{\mu\nu} = f_{s0}\mu_1/2 + a\nu_1$, $f_{\nu\mu} = f_{s0}\nu_1/2 + a\mu_1$, $\mu_j = \cosh(p_jd_j)$, $\nu_j = \sinh(p_jd_j)$, $\eta_j = \cosh(q_jd_j)$, $\zeta_j = \sinh(q_jd_j)$, $Q_j = \tanh(q_jd_j)$, and $P_j = \tanh(p_jd_j)$. For compactness below we provide explicit expressions for the matching constants in the case of equal normal-state conductivities $\sigma_{F_1} = \sigma_{F_2}$ and generalize the results for $\sigma_{F_1}/\sigma_{F_2} \neq 1$ afterwards.
\begin{widetext}
We obtain
\begin{subequations}
\begin{align}
c = \frac{if_{s0}}{\eta_2R(\theta)}\sin(\theta)\left[\cos(\theta)W^{(c)}_1 + W^{(c)}_2\right] \ ,\\
b = \frac{if_{s0}}{\eta_1R(\theta)}\sin(\theta)\left[\cos(\theta)W^{(b)}_{1} + W^{(b)}_{2}\right] \ ,
\end{align}
\end{subequations}
where
\begin{subequations}
\begin{align}
W_1^{(c)} = {\rm Im}(p_2P_2)\left[|p_1|^2Q_2{\rm Re}(\mu_1^{-1})-q_1{\rm Re}(p_1P^*_1\mu_1^{-1})\right] \ ,\\
W_2^{(c)} = {\rm Im}(\mu_1^{-1})|p_1|^2\left[q_1 + Q_1{\rm Re}(p_2P_2)\right] + 
 {\rm Im}(p_1P^*_1\mu_{1}^{-1})\left[Q_1|p_2P_2|^2 + q_1{\rm Re}(p_2P_2)\right] \ ,\\
W_1^{(b)} = {\rm Im}(\mu_1^{-1})|p_1|^2\left[{\rm Re}(p_2P_2)-q_2Q_2\right] +
 {\rm Im}(p_1P_1^*\mu_1^{-1})\left[|p_2P_2|^2 - q_2Q_2{\rm Re}(p_2P_2)\right] \ ,\\
W_2^{(b)} = {\rm Im}(p_2P_2)\left[|p_1|^2{\rm Re}(\mu_{1}^{-1}) + q_2Q_2{\rm Re}(p_1P^*_1\mu_1^{-1})\right] \ .
\end{align}
\end{subequations}
The remaining coefficients have the form
\begin{subequations}
\begin{align}
c_+ = \frac{f_{s0}}{4\mu_2R(\theta)}\left[W_1^{(c_+)} + \cos(\theta)W_2^{(c_+)} + \cos(2\theta)W_3^{(c_+)}\right] \ ,\\
a = \frac{f_{s0}}{4R(\theta)}\left[W_1^{(a)} + \cos(\theta)W_2^{(a)} + \sin^2(\theta)W_3^{(a)}\right] \ ,
\end{align}
\end{subequations}
with
\begin{subequations}
\begin{align}
W_1^{(c_+)} = {\rm Re}(\mu_1^{-1})|p_1|^2\left[2q_1 + q_2Q_1Q_2 + Q_1p^*_2P^*_2\right] + 
 {\rm Re}(p_1P^*_1\mu_1^{-1})\left[p^*_2P^*_2(q_1 + 2q_2Q_1Q_2) + q_1q_2Q_2\right] \ ,\\
W_2^{(c_+)} = 2i(q_1 + q_2Q_1Q_2)\left[|p_1|^2{\rm Im}(\mu_1^{-1}) - p^*_2P^*_2{\rm Im}(p_1P^*_1\mu_1^{-1})\right] \ ,\\
W_3^{(c_+)} = 2(q_2Q_2 - p_2^*P^*_2)\left[|p_1|^2Q_2{\rm Re}(\mu_1^{-1}) - q_1{\rm Re}(p_1P^*_1\mu_1^{-1})\right] \ ,\\
W_1^{(a)} = -2(q_1 + q_2Q_1Q_2)\left[|p_1|^2P_1 + |p_2P_2|^2P^*_1 + {\rm Re}(p_2P_2)(p^*_1 + p_1 |P_1|^2)\right] \ ,\\
W_2^{(a)} = -2i(q_1 + q_2Q_1Q_2){\rm Im}(p_2P_2)(p^*_1 - p_1|P_1|^2) \ ,\\
W_3^{(a)} = {\rm Re}(p_2P_2)\bigl[p^*_1q_2Q_1Q_2(1 + |\mu_1|^{-2}) + p_1^*q_1(1 - |\mu_1|^{-2}) \\
\nonumber
-2|p_1|^2P_1Q_1 + p_1|P_1|^2(q_2Q_1Q_2 + q_1)-2q_1q_2Q_2P^*_1\bigl] + 
+2|p_1|^2q_2P_1Q_1Q_2 + 2|p_2P_2|^2q_1P^*_1 - \\
\nonumber
 q_1q_2Q_2\left[p^*_1(1 - |\mu_1|^{-2}) + p_1|P_1|^2\right] - Q_1|p_2P_2|^2\left[p^*_1(1 + |\mu_1|^{-2}) + p_1\right] \ .
\end{align}
\end{subequations}
In the above expressions 
\begin{subequations}
\begin{align}
R(\theta) = T_0 + T_1\cos(\theta) + T_2\sin^2(\theta) \ ,\\
T_0 = (q_1 + q_2Q_1Q_2)\left[|p_1|^2 + |p_2P_1P_2|^2 + 2{\rm Re}(p_2P_2){\rm Re}(p_1P^*_1)\right] \ ,\\
T_1 = 2(q_1 + q_2Q_1Q_2){\rm Im}(p_1P^*_1){\rm Im}(p_2P_2) \ ,\\
T_2 = {\rm Re}(p_2P_2)\left[-(q_1 + q_2Q_1Q_2){\rm Re}(p_1P^*_1) + |p_1|^2Q_1 + |P_1|^2q_1q_2Q_2\right] - \\
\nonumber
-|p_1|^2q_2Q_1Q_2 - q_1|p_2P_1P_2|^2 + {\rm Re}(p_1P^*_1)(q_1q_2Q_2 + |p_2P_2|^2Q_1) \ .
\end{align}
\end{subequations}
For the conductivity ratio $\sigma_{F_2}/\sigma_{F_1}$ different from unity one should replace $p_2 \to p_2\sigma_{F_2}/\sigma_{F_1}$ and $q_2\to q_2\sigma_{F_2}/\sigma_{F_1}$ in the above expressions whereas the products $p_2d_2$ and $q_2d_2$ are unchanged. 
\end{widetext}

\section{Derivation of Eq.~(\ref{magnetic_free_energy}) in the main text}\label{App_C}
Here we provide the derivation of the magnetic free energy per unit cross-sectional area presented by Eq.~(\ref{magnetic_free_energy}) in the main text.  Assuming the thicknesses of the ferromagnets to be much smaller than the London penetration depth, the solution of the Maxwell equation~(\ref{Maxwell_equation}) can be written in the form 
\begin{subequations}
\begin{align}
x<0: \ {\bf A}&={\bf A}_0\exp\left(x/\lambda_0\right) \ ,\\
0<x<d_1: \  {\bf A}&={\bf A}_0+4\pi x {\bf M}_1\times{\bf e}_x  \ ,\\
d_1<x<d_1+d_2: \  {\bf A}&={\bf A}_0+
4\pi d_1 {\bf M}_1\times{\bf e}_x  \\
\nonumber
+
4\pi \left(x-d_1\right){\bf M}_2\times{\bf e}_x \ ,
\end{align}
\end{subequations}
where the vector ${\bf A}_0$ lays in the $yz$-plane and $\lambda_0$ is the London penetration depth inside the superconductor. Note that for a certain vector ${\bf A}(x)={\bf a}_0f(x)$ laying in the $yz$-plane one gets ${\rm curl}{\bf A}=\left({\bf e}_x\times{\bf a}_0\right)\left(\partial f/\partial x\right)$. Then the magnetic field ${\bf B}={\rm curl}{\bf A}$ inside the superconductor has the form
\begin{equation}
{\bf B}=\frac{1}{\lambda_0}\left({\bf e}_x\times{\bf A}_0\right)\exp\left(x/\lambda_0\right).
\end{equation}
Integrating the Maxwell equation~(\ref{Maxwell_equation}) across the ferromagnetic bilayer F$_1$/F$_2$ and taking into account the London relation between the superconducting current and the vector potential $\mathbf{j}_s(x) = -(c/4\pi)\lambda^{-2}(x)\mathbf{A}(x)$, we find
\begin{eqnarray}\label{A0_implicit_expression}
-\frac{1}{\lambda_0}{\bf A}_0={\bf A}_0\int\limits_{0}^{d_1}\frac{dx}{\lambda^2(x)} + \mathbf{m}_1\int\limits_{0}^{d_1}\frac{xdx}{\lambda^2(x)} + \\
\nonumber
\left({\bf A}_0+\mathbf{m}_1d_1\right)\int\limits_{d_1}^{d_1+d_2}\frac{dx}{\lambda^2(x)}+\mathbf{m}_2\int\limits_{d_1}^{d_1+d_2}\frac{\left(x-d_1\right)dx}{\lambda^2(x)}.
\end{eqnarray}
Here $\mathbf{m}_j = 4\pi (\mathbf{M}_j\times\mathbf{e}_x)$ and $j = 1,2$. From the above equation one may find the vector constant ${\bf A}_0$. In the case of a large mismatch between the normal-state conductivities $\sigma_S\gg \sigma_{F_j}$ ($j = 1,2$), where $\sigma_S$ ($\sigma_{F_j}$) is the normal-state conductivity in the S (F$_j$) layer, one can neglect the terms $\propto \mathbf{A}_0$ in the right-hand side of Eq.~(\ref{A0_implicit_expression}) due to the fact that $\lambda_0/\lambda \propto \sqrt{\sigma_{F_j}/\sigma_S} \ll 1$. As a result, we obtain
\begin{subequations}
\begin{align}\label{A0_conductivity_mismatch}
\mathbf{A}_0 = -\lambda_0\left[\mathbf{m}_1(Q_1^{(1)} + d_1Q_0^{(2)}) + \mathbf{m}_2Q_1^{(2)}\right] \ ,\\
Q_n^{(1)}=\int\limits_{0}^{d_1}\frac{x^ndx}{\lambda^2(x)} \ , \ \ \ Q_n^{(2)}=\int\limits_{d_1}^{d_1+d_2}\frac{\left(x-d_1\right)^ndx}{\lambda^2(x)} \ .
\end{align}
\end{subequations}
Note that the terms appearing in the right-hand side of Eq.~(\ref{A0_conductivity_mismatch}) are of the order of $\lambda_0M_1(d_1/\lambda)^2$, $\lambda_0M_1d_1d_2/\lambda^2$, and $\lambda_0 M_2(d_2/\lambda)^2$, respectively. We proceed with calculations of the magnetic part of the free energy density per unit cross sectional area 
\begin{equation}\label{F_def}
F_m=\frac{1}{8\pi}\int\limits_{-\infty}^{d_1+d_2}\left[\left({\rm rot}{\bf A}-4\pi {\bf M}\right)^2+ \lambda^{-2}{\bf A}^2\right]dx.
\end{equation}
It is convenient to consider the contributions to the magnetic energy from each layer separately and introduce $\tilde{F}_m=8\pi F_m$, so that $\tilde{F}_m = \tilde{F}_{m}^{(S)} + \tilde{F}_{m}^{(F_{1})}  + \tilde{F}_{m}^{(F_2)}$. Here the upper subscripts refer to the corresponding layers. It follows from the above estimates of $\mathbf{A}_0$ that the free energy corresponding to the superconductor 
\begin{equation}\label{Fs}
\tilde{F}_m^{(S)}=\int\limits_{-\infty}^{0}\left({\bf B}^2+ \lambda_0^{-2}{\bf A}^2\right)dx={\bf A}_0^2/\lambda_0 \ ,
\end{equation}
vanishes in the limit $\sigma_{F_j}/\sigma_S \to 0$. Therefore, the relevant contributions to the magnetic energy originate only from ferromagnetic layers and can be written as follows:
\begin{subequations}\label{magnetic_energy_contributions_ferromagnets}
\begin{align}
\tilde F_m^{(F_1)}=\int\limits_{0}^{d_1}\lambda^{-2}{\bf A}^2dx= \\
\nonumber
{\bf A}_0^2Q_0^{(1)}+2\left({\bf A}_0\cdot{\bf m}_1\right)Q_1^{(1)}+{\bf m}_1^2Q_2^{(1)} \ ,\\
\tilde F_m^{(F_2)}=\int\limits_{d_1}^{d_1+d_2} \lambda^{-2}{\bf A}^2dx =\left({\bf A}_0+{\bf m}_1d_1\right)^2Q_0^{(2)}+\\
\nonumber
2\left({\bf A}_0+{\bf m}_1d_1\right)\cdot{\bf m}_2Q_1^{(2)}+{\bf m}_2^2Q_2^{(2)}.
\end{align}
\end{subequations}
Since $\mathbf{A}_0$ has the magnitude of the order of $M\lambda_0(d/\lambda)^2$ with $M = \max\{M_1,M_2\}$ and $d = \max \{d_1,d_2\}$, the contributions proportional to $\mathbf{A}_0$ and $\mathbf{A}_0^2$ in the above expressions vanish in the limit $\sigma_{F_j}/\sigma_S \to 0$ (or $\lambda_0/\lambda \to 0$). Indeed, comparing different terms in Eq.~(\ref{magnetic_energy_contributions_ferromagnets}) we find that $|\mathbf{A}_0|/|\mathbf{m}_1|d_1 \sim \lambda_0d/\lambda^2$, $\mathbf{A}_0^2Q_0^{(1)}/\mathbf{m}_1^2Q_2^{(1)}\sim \lambda_0^2 d^2/\lambda^4$, and $\mathbf{A}_0\mathbf{m}_1 Q_1^{(1)}/\mathbf{m}_1^2Q_2^{(1)}\sim (\lambda_0 d/\lambda^2)$. Thus, in the case of a large mismatch between the normal-state conductivities the magnetic free energy density per unit cross sectional area can be presented in the form 
\begin{eqnarray}\label{F_res}
\tilde F={\bf m}_1^2\left(Q_2^{(1)}+d_1^2Q_0^{(2)}\right)+2\left({\bf m}_1\cdot{\bf m}_2\right)d_1Q_1^{(2)}+\\
\nonumber
{\bf m}_2^2Q_2^{(2)}+O\left[\left(d/\lambda\right)^4\right].
\end{eqnarray}
Note that all these terms are not sensitive to the geometry of the superconductor. In fact, the superconductor here is just a generator of Cooper pairs. The resulting Eq.~(\ref{F_res}) can be rewritten in the form presented by Eq.~(\ref{magnetic_free_energy}) in the main text.

\end{document}